\documentclass[preprint]{amsart}
\usepackage{amsmath}
\usepackage{mathrsfs}
\usepackage{amssymb}
\usepackage{mathtools}
\usepackage[dvipdfmx]{color}

\def\dimm{m}
\def\dimg{f}
\def\dimb{b}
\def\dimw{w}
\def\dimrw{{\widetilde w}}
\def\kili{I}
\def\kilj{J}
\def\kilk{K}
\def\kill{L}

\def\homi{i}
\def\homj{j}
\def\homk{k}

\def\orbm{\mu}
\def\orbn{\nu}
\def\orbl{\lambda}
\def\orbr{\rho}
\def\orba{\alpha}

\def\wsA{A}
\def\wsB{B}
\def\wsC{C}
\def\wsD{D}
\def\sta{a}
\def\stb{b}
\def\stc{c}
\def\std{d}

\def\wsi{i}

\def\redA{{A'}}
\def\redB{{B'}}
\def\redC{{C'}}

\title{Note on orbit space of $G$ membranes}
\author{Mitsuharu Hasegawa and Daisuke Ida}
\address{Department of Physics, Gakushuin University, Tokyo 171-8588}
\date{21 September 2018}

\begin{document}
\maketitle

\begin{abstract}
The motion of test membranes on which the group $G$ of isometries of
  a spacetime $M$
  acts has been considered in general settings.
  It has been shown that the configuration of Nambu-Goto membranes is
  described by the Nambu-Goto membranes in the quotient manifold $M/G$
  with an appropriate projected metric
  if (i) $G$ is Abelian, (ii) $G$ is semisimple and compact,
  or (iii) the orthogonal distribution of the orbit of $G$ is integrable,
  but in general not.

  It has also been shown that a similar result holds when the membranes couple
  with scalar maps or differential form fields.
  
  \end{abstract}

\section{Introduction}
Extended objects in cosmology such as
cosmic strings and membranes (domain walls) 
naturally arise as topological defects
associated with various symmetry breaking phenomena in
quantum field theory.
They play an important role in the 
scenario of the structure formation of the Universe.

The description of motion of extended objects such as strings and membranes
in a spacetime with Killing vector fields
simplifies when they respect the spacetime symmetry.

This kind of simplification has long been known in
the minimal surface theory~\cite{HL71} as the cohomogeneity technique.
In general relativity, Frolov {\it et al.}~\cite{FSZH89} find that the configuration
of stationary Nambu-Goto strings in a stationary spacetime is determined
by the geodesic equation in a certain Riemannian 3-manifold.
Their reasoning is based on the
observation that when the  stationary string ansatz
is substituted into the Nambu-Goto action, the action
reduces to the geodesic action via the dimensional reduction.

This technique is widely applied to the construction of stationary
string solutions in various background spacetimes~\cite{CF89,SS89,CFL98,FLC99,FS04,AA08}.
It is also useful to find dynamical string solutions in spacetimes with
a Killing vector field~\cite{IK05,KKI08,KKI10,II10a,II10b}.

A similar idea for this dimensional reduction also works
for $\dimg+1$-dimensional Nambu-Goto membranes when
the spacetime has $\dimg$  
pairwise commuting 
Killing vector fields
and the membrane respects this symmetry~\cite{KF08}.

The authors of Ref.~\cite{KKI15} pointed out that
the dimensional reduction at the action level also occurs 
when the $\dimg$ Killing vector fields are noncommuting.
With this observation, they claim that the $\dimg+1$-dimensional 
Nambu-Goto membranes respecting the $\dimg$-dimensional 
non-Abelian group of isometry of the spacetime could be
reduced to the geodesic motion in the quotient manifold.

  However, it is of course not the correct procedure
to put a trial solution directly into the action.
Hence, we would like to confirm whether the above claim is correct or not.

In the following, we consider Nambu-Goto membranes of general dimensions
in spacetime with a non-Abelian group of isometries, 
assuming that the membranes respect the spacetime symmetry. 
We show that the resultant equation of motion for
the membranes is almost that for lower-dimensional Nambu-Goto membranes, but
with extra force terms. This force term becomes zero for the Abelian case,
or the semisimple and compact cases, but it does not in general.
In particular, the claim in Ref.~\cite{KKI15} is not the case.

The organization of this paper is as follows.
In Sec. \ref{sec:isom}, the mathematical settings are described.
In Sec. \ref{sec:g-inv}, the Nambu-Goto membranes with spacetime symmetries are
considered, and their general equation of motion is derived.
In Sec. \ref{sec:4}, similar consideration on 
the membranes coupled with scalar maps is made.
In Sec. \ref{sec:5}, membranes coupled with a differential form field are treated.
In Sec. \ref{sec:6}, several remarks are made.

\section{Isometric actions  on world sheets}
\label{sec:isom}

Let $(M,g)$ be an $\dimm$-dimensional spacetime,
which is a differentiable manifold $M$ endowed with
a spacetime metric $g$ of the signature $(-,+,\dots,+)$.
Let  $G$ be an $\dimg$-dimensional connected Lie subgroup
of the full isometry group of $M$.

The  (left) $G$-action  on $M$ is a group homomorphism
\begin{align*}
G\longrightarrow \operatorname{Diff}(M);~~~
g\longmapsto f_g
\end{align*}
of $G$ into the group of diffeomorphisms on $M$,
such that 
\begin{align*}
G\times M\longrightarrow M;~~~
(g,x)\longmapsto f_g(x) 
\end{align*}
is differentiable.

We assume that the $G$-action on $M$ is  free,
which means that $f_g$ does not have a fixed point on $M$ for every nonidentity element $g$
of $G$, or in other words, $M$ admits $\dimg$ linearly independent Killing vector fields as the infinitesimal generators of $G$.

We also require that the $G$-action on  $M$ be  proper, i.e.,
the map
\begin{align*}
G\times M\longrightarrow M\times M;
(g,x)\longmapsto (f_g(x),x)
\end{align*}
is proper, which means that the preimage of any compact set is compact.

For each element $x$ in $M$,
the set
\begin{align*}
Gx=\{f_g(x)|g\in G\}
\end{align*}
is called the orbit of $x$.
The set of these orbits is called the  orbit space, and it is denoted by
$M/G$.

Under the free and proper action of $G$, it is guaranteed that
(a) each orbit $Gx$ is an embedded closed submanifold of $M$, 
(b) $Gx$ is diffeomorphic with $G$, 
(c) the  orbit space $M/G$ naturally acquires a differentiable
structure.

In the rest of this section, 
we give the general form of the Lorentzian metric $g$
on $M$. The construction goes along  similar lines to that of
homogeneous universes \cite{EM68}.

We first determine the geometry of the orbit $Gx$.
We assume that each orbit is non-null, so that
the induced metric on $Gx$ is a nondegenerate Riemannian or Lorentzian metric.

Let $\{y^\homi\}_{\homi=1,2,\dots,\dimg}$ be a local coordinate system on $Gx$.
Since $\dimg$ linearly independent Killing vector fields 
$\{\xi_\kili\}_{\kili=1,2,\dots,\dimg}$ generating a $G$-action
are tangent to $Gx$, these can be written as
$\xi_\kili=\xi_{\kili}{}^\homi \partial_\homi$  with this coordinate basis.

Since these Killing vector fields generate a left action of $G$ on $Gx$, 
  they
are identified with the right invariant vector fields on $G$.
According to the general theory of Lie groups, the right invariant
vector fields on $G$ are subject to the commuting relation
\begin{align*}
  [\xi_\kili,\xi_\kilj]=f_{\kili\kilj}{}^\kilk\xi_\kilk,
\end{align*}
where $f_{\kili\kilj}{}^\kilk$'s are the structure constants for
the Lie algebra $\frak{g}$ of $G$.

A left invariant vector field $\sigma^\homi$ on $Gx$ is a tangent vector field
invariant under the $G$-action, characterized by the equation
\begin{align*}
  \mathscr{L}_{\xi_\kili}\sigma^\homi=0,
\end{align*}
where $\mathscr{L}_{\xi_\kili}$ denotes the Lie derivative.
This equation admits $\dimg$ linearly independent solutions,
which we denote by $\{\sigma_\kili{}^\homi\}_{\kili=1,2,\dots,\dimg}$.
By taking a linear combination,
  it is always possible to find
the basis $\{\sigma_I^i\}_{I=1,2,\dots,f}$
of the left invariant vector fields, such  that
\begin{align*}
  [\sigma_{\kili},\sigma_{\kilj}]=f_{\kili\kilj}{}^\kilk\sigma_\kilk
\end{align*}
holds.

The dual basis of 1-forms $\{\sigma^\kili{}_\homi\}_{\kili=1,2,\dots,\dimg}$,
  characterized by
\begin{align*}
  \sigma_\kili{}^\homk\sigma^\kilj{}_\homk=\delta^\kilj_\kili,
\end{align*}
consists of left invariant 1-forms.
  These satisfy
\begin{align*}
  \mathscr{L}_{\xi_\kili}\sigma^\kilj{}_\homk&=0,\\
  d\sigma^\kilk&=-\dfrac{1}{2}f_{\kili\kilj}{}^\kilk\sigma^\kili\wedge\sigma^\kilj.
\end{align*}

The induced metric on $Gx$ can be written in terms of this left invariant
basis as
\begin{align}\label{eq:2-1}
  g_{\homi\homj}=\phi_{\kili\kilj}\sigma^\kili{}_\homi\sigma^\kilj{}_\homj,
\end{align}
with entries  $\phi_{\kili\kilj}$ of
the nondegenerate symmetric matrix.
Since $g_{\homi\homj}$ is invariant under the $G$-action, i.e.,
\begin{align*}
  \mathscr{L}_{\xi_\kili}g_{\homi\homj}=0
\end{align*}
should be required,
the coefficients
$\phi_{\kili\kilj}$ are constants over $Gx$.
  The Equation (\ref{eq:2-1})
  gives the general form of the metric  on the orbit.

Now, we can write the spacetime metric $g$ in the present setting.
Since the orbit space $M/G$ is the differentiable manifold,
it has a local coordinate system, which we denote by
$\{z^\orbm\}_{\orbm=1,2,\dots,\dimb}$, where $\dimb=\dimm-\dimg$.
  The spacetime metric could  in general be written as
\begin{align*}
  g&=\phi_{\kili\kilj}(z^\orba)
\left(\sigma^\kili(y^\homk)-w^\kili{}_\orbm(y^\homk,z^\orbl) dz^\orbm\right)
\left(\sigma^\kilj(y^\homk)-w^\kilj{}_\orbn(y^\homk,z^\orbl) dz^\orbn\right)\\
&+h_{\orbm\orbn}(y^\homk,z^\orbl)dz^\orbm dz^\orbn,
\end{align*}
  so that it induces
$g_{\homi\homj}=\phi_{\kili\kilj}\sigma^\kili{}_\homi\sigma^\kilj{}_\homj$
  on each orbit by setting $z^\orbm={\rm const}$.
  Since the spacetime metric $g$ is invariant under the $G$-action,
  it is subject to
the Killing equation
\begin{align*}
  \mathscr{L}_{\xi_\kili}g=0.
\end{align*}
Then, it is required that
\begin{align*}
w^\kili{}_\orbn=w^\kili{}_\orbn(z^\homk),~~
h_{\orbm\orbn}=h_{\orbm\orbn}(z^\homk)
\end{align*}
hold.
This gives the general local form of the spacetime metric $g$.

\section{Nambu-Goto $G$ membranes}
\label{sec:g-inv}
Let us consider the motion of extended objects in spacetimes equipped
with the isometric $G$-action.
It is generally expected that the equation of motion simplifies when
the extended object also respects the isometry.
The simplification typically occurs in the form of the dimensional reduction;
i.e., the equation of motion reduces to that for the objects
(e.g., particles, strings, or membranes) in the orbit space
$M/G$.
In this section,  we study the Nambu-Goto membranes  as
a basic example of extended objects in the relativistic mechanics.

In general, a relativistic membrane is described as 
a timelike immersion
$
i:W\longrightarrow M
$
of a differentiable manifold $W$, called a world sheet, into the spacetime $M$.
Let $\{x^\sta\}_{\sta=1,2,\dots,\dimm}$ 
and  $\{s^\wsA\}_{\wsA=1,2,\dots,\dimw}$ be a local coordinate system on $M$
and $W$, respectively.
  The immersion $i$ is locally described as
\begin{align*}
  x^\sta=X^\sta(s^1,\dots,s^\dimw),
\end{align*}
in terms of the $\dimm$ scalar functions $X^\sta$ on the world sheet $W$.
Then, the Lorentzian metric
\begin{align*}
  G_{\wsA\wsB}=g_{\sta\stb}X^\sta{}_{,\wsA}X^\stb{}_{,\wsB}
\end{align*}
is locally induced on $W$.

The Nambu-Goto action for the relativistic membrane
is given by 
\begin{align*}
  S[X^\sta]=-\tau\int_W ds^1\dots ds^\dimw
\sqrt{|G|},
\end{align*}
where $\tau$ is a constant 
  identified with
the tension of the membrane,
and  $G=\operatorname{det}G_{\wsA\wsB}$.
The Euler-Lagrange equation becomes
\begin{align*}
K^\sta:= D_\wsC D^\wsC X^\sta+\Gamma^\sta_{\stb\stc}\mathscr{G}^{\stb\stc}=0,
\end{align*}
where $D_\wsC$ denotes the covariant derivative 
with respect to $G_{\wsA\wsB}$,
$\Gamma^\sta_{\stb\stc}$
the Christoffel symbol 
for  $g_{\sta\stb}$, and
\begin{align*}
  \mathscr{G}^{\sta\stb}=G^{\wsA\wsB}(D_{\wsA}X^\sta)D_{\wsB}X^\stb
\end{align*}
has been defined, where $G^{\wsA\wsB}$ denotes the inverse matrix of $G_{\wsA\wsB}$.

The vector field $K^\sta$ on $W$ has a simple geometrical meaning.
It is the mean curvature vector
\begin{align*}
  K^a=G^{\wsB\wsC}K^a_{\wsB\wsC},
\end{align*}
  that is
the trace of the extrinsic curvature vector
\begin{align*}
  K^\sta_{\wsB\wsC}=D_\wsB D_\wsC X^\sta+\Gamma^a_{bc}(D_\wsB X^b)D_\wsC X^c.
\end{align*}
  The extrinsic curvature vector  is defined as follows:
 let $U^\wsA$ and $V^\wsA$ be tangent vector fields
on $W$, and let $\mathscr{U}^\sta$ and $\mathscr{V}^\sta$ be
smoothly extended vector fields
of $i_*U$ and $i_*V$, respectively,
to a neighborhood $\mathscr{W}$ of
 $W$. 
  For $x\in W$, the orthogonal decomposition
 $T_x M=T_x W\oplus N_x W$ of $(\nabla_{\mathscr{U}}\mathscr{V})_x$
  is written as
 \begin{align*}
\nabla_{\mathscr{U}}\mathscr{V}=D_U V+K(U,V). 
 \end{align*}
 Then, $K:T_xW\times T_x W\longrightarrow N_x W$ is defined by
this equation.

We assume that the group of isometry $G$ acts freely and properly on $M$.
Then, as we see in Sec.~\ref{sec:isom}, the spacetime metric can be written
locally as
\begin{align*}
  g=\phi_{\kili\kilj}
\left(\sigma^\kili-w^\kili{}_\orbm dz^\orbm\right)
\left(\sigma^\kilj-w^\kilj{}_\orbn dz^\orbn\right)
+h_{\orbm\orbn}dz^\orbm dz^\orbn,
\end{align*}
where
$\sigma^\kili=\sigma^\kili(y^\homk)$ constitutes a left invariant basis 
of $1$-forms on the orbit $Gx$, and
\begin{align*}
  \phi_{\kili\kilj}=  \phi_{\kili\kilj}(z^\orbl),~~
w^\kili{}_\orbm =w^\kili{}_\orbm (z^\orbl),~~
h_{\orbm\orbn}=h_{\orbm\orbn}(z^\orbl)
\end{align*}
are the scalar, vector and metric tensor fields on the orbit space
$M/G$.

Let the membrane respect this isometry, so that
the image of the world sheet $W$ is $G$-invariant, 
i.e. invariant under the action of $G$.
The general form of such $G$ membranes can be  written  as
\begin{align*}
\begin{array}{ll}
  X^\homi=\alpha^\homi&
(i=1,\dots,\dimg)\\
X^\orbm=X^\orbm(\beta^\redA)&
(\orbm=\dimg+1,\dots,\dimm)
\end{array}
\end{align*}
in terms of the world sheet coordinates 
\begin{align*}
\{s^\wsA\}=
\{\alpha^\wsi;\beta^\redA\}_{\wsi=1,\dots,\dimg,\redA=\dimg+1,\dots\dimw}.
\end{align*}
This is identified with the immersion of the world sheet orbit space
$W/G$ into the spacetime orbit space $M/G$, characterized by
\begin{align*}
  X^\orbm=X^\orbm(\beta^\redA).
\end{align*}
Thus, a $G$ membrane can be regarded as a membrane in the orbit space
$M/G$.

Although the following calculations are most efficiently executed
via the Cartan's structure equations for connection forms,
we show the results of the direct coordinate calculations for the reader's
convenience.
In the following calculations, indices are raised and lowered,
respectively,
by $\phi^{\kili\kilj}$, $\phi_{\kili\kilj}$, $\lambda^{\homi\homj}$, 
$\lambda_{\homi\homj}$,
$h^{\orbm\orbn}$, and $h_{\orbm\orbn}$, where $\phi^{\kili\kilj}$ denotes
the entries of the inverse matrix of $\phi_{\kili\kilj}$,
$\lambda_{\homi\homj}$ is defined by
\begin{align*}
  \lambda_{\homi\homj}=\phi_{\kili\kilj}\sigma^\kili{}_{\homi}\sigma^\kilj{}_{\homj},
\end{align*}
$\lambda^{\homi\homj}$ is its inverse, and $h^{\orbm\orbn}$ is the inverse 
of $h_{\orbm\orbn}$.
The covariant derivative compatible with $h_{\orbm\orbn}$ is denoted by
the semicolon.

The components of the spacetime metric $g$ and its inverse $g^{-1}$
are given by
\begin{align*}
  g_{\homi\homj}&=\lambda_{\homi\homj},~~~
g_{\homi\orbm}=-w_{\homi\orbm},~~~
g_{\orbm\orbn}=h_{\orbm\orbn}+w_{\kilk\orbm}w^\kilk{}_\orbn,\\
  g^{\homi\homj}&
=\lambda^{\homi\homj}
+\sigma_\kili{}^\homi\sigma_\kilj{}^\homj w^\kili{}_\orbl w^{\kilj\orbl},~~~
g^{\homi\orbm}=\sigma_\kilk{}^\homi w^{\kilk\orbm},~~~
g^{\orbm\orbn}=h^{\orbm\orbn}.
\end{align*}
The Christoffel symbols are computed as
\begin{align}
\label{eq:gamijk}  \Gamma^\homi_{\homj \homk }&=\sigma_\kili{}^\homi\sigma^\kili{}_{(\homj,\homk )}
+\sigma_\kili{}^\homi\sigma^\kilj {}_\homj\sigma^\kilk {}_\homk 
\left[f^\kili{}_{(\kilj \kilk )}+f_{\kill(\kilj \kilk )}w^\kili{}_\orbm w^{\kill \orbm}
-\dfrac{1}{2}\phi_{\kilj \kilk ,\orbm}w^{\kili \orbm}\right],\\
\label{eq:gamijlam} 
\Gamma^\homi_{\homj \orbl }
&=\sigma_\kili {}^\homi\sigma^\kilj {}_\homj \biggl[
\dfrac{1}{2}\phi_{\kilj \kilk ,\orbl }\phi^{\kili \kilk }
+\dfrac{1}{2}\phi_{\kilj \kilk ,\orbr }w^{\kili \orbr }w^\kilk {}_\orbl 
\\
\nonumber
&-\dfrac{1}{2}f^\kili {}_{\kilj \kilk }w^\kilk {}_\orbl 
-\dfrac{1}{2}f_{\kill \kilj \kilk }w^{\kili \orbr }w^{\kill }{}_{\orbr }w^\kilk {}_\orbl 
-w^{\kili \orbr }\phi_{\kilj \kilk }w^\kilk {}_{[\orbr ;\orbl ]}
\biggr],\\
\label{eq:gaminulam} 
\Gamma^\homi_{\orbn \orbl }&=\sigma_\kilk {}^\homi\biggl[
-\phi^{\kili \kilk }\phi_{\kili \kilj ,(\orbn }w^\kilj {}_{\orbl )}
-w^\kilk {}_{(\orbn ,\orbl )}
+w^\kilk {}_\orbm{}^h\Gamma^\orbm_{\orbn \orbl }\\
\nonumber&+w^{\kilk \orbm}(
-\dfrac{1}{2}\phi_{\kili \kilj ,\orbm}w^\kili {}_\orbn w^\kilj {}_\orbl 
+w_{\kili \orbn }w^\kili {}_{[\orbm;\orbl ]}
+w_{\kili \orbl }w^\kili {}_{[\orbm;\orbn ]}
)
\biggr],\\
\label{eq:gammujk} 
\Gamma^\orbm_{\homj \homk }&=
\sigma^\kilj {}_\homj \sigma^\kilk {}_\homk 
\left(-\dfrac{1}{2}\phi_{\kilj \kilk }{}^{,\orbm}
+f_{\kili (\kilj \kilk )}w^{\kili \orbm}\right),\\
\label{eq:gammunuk} 
\Gamma^\orbm_{\orbn \homk }&=
\sigma^\kilk {}_\homk \left(
\dfrac{1}{2}\phi_{\kilj \kilk }{}^{,\orbm} w^\kilj {}_\orbn 
+\phi_{\kilj \kilk }h^{\orbm \orbl }w^\kilj {}_{[\orbn ;\orbl ]}
-\dfrac{1}{2}f_{\kili \kilk \kilj }w^{\kili \orbm }w^\kilj {}_\orbn 
\right),\\
\label{eq:gammunulam} 
\Gamma^\orbm _{\orbn \orbl }&=
{}^h\Gamma^\orbm _{\orbn \orbl }
-\dfrac{1}{2}\phi_{\kili \kilj }{}^{,\orbm }w^\kili {}_\orbn w^\kilj {}_\orbl 
+w_{\kili (\orbn }w^{\kili \orbm }{}_{;\orbl )}
-w_{\kili (\orbn }w^\kili {}_{\orbl )}{}^{;\orbm },
\end{align}
where ${}^h\Gamma^\orbm_{\orbn\orbl}$ denotes the Christoffel symbol
with respect to $h_{\orbm\orbn}$.
Note that we raise or lower the indices $\kili,\kilj,\dots$ with
$\phi_{\kili\kilj}$, $\phi^{\kili\kilj}$,
but not with the Killing metric on the Lie algebra $\frak{g}$, so that,
e.g., $f_{\kili\kilj\kilk}$ may not be totally antisymmetric under the
permutation of the indices.

The induced metric $G_{\wsA\wsB}$ and its inverse
$G^{\wsA\wsB}$ on the world sheet $W$ become
\begin{align*}
G_{\homi \homj }&=\lambda_{\homi \homj },~~
G_{\homi \redB}=-\lambda_{\homi \homj }C^\homj {}_{\redB},~~
G_{\redA \redB}=G'_{\redA \redB}
+\lambda_{\homi \homj }C^\homi {}_{\redA }C^\homj {}_{\redB}\\
G^{\homi \homj }&=\lambda^{\homi \homj }+C^\homi {}_{\redA }C^\homj {}_{\redB}G'^{\redA \redB},~~
G^{\homi \redB}=C^\homi {}_{\redA }G'^{\redA \redB},~~
G^{\redA \redB}=G'^{\redA \redB}
\end{align*}
where
\begin{align*}
C^\homj {}_{\redB}&=\sigma_\kilj{}^\homj w^\kilj{}_\orbl D_{\redB}X^\orbl,\\
 G'_{\redA \redB}&=h_{\orbm \orbn}(D_{\redA}X^\orbm)D_{\redB}X^\orbn
\end{align*}
have been defined, and $G'^{\redA\redB}$ denotes the inverse of
$G'_{\redA\redB}$.
This $G'_{\redA \redB}$ gives the induced metric on the quotient world sheet
$W/G$ as the membrane immersed in $(M/G,h)$.

The spacetime component
$\mathscr{G}^{\orbm\orbn}$
of $G^{\wsA\wsB}$ is calculated as
\begin{align*}
\mathscr{G}^{\homi \homj }&
=\sigma_\kili{}^\homi\sigma_\kilj{}^\homj(\phi^{\kili\kilj}+w^\kili{}_\orbm w^\kilj{}_\orbn \mathscr{G}'^{\orbm\orbn}),\\
\mathscr{G}^{\homi \orbn}&=
\sigma_\kili {}^\homi w^\kili {}_\orbm\mathscr{G}'^{\orbm\orbn},\\
\mathscr{G}^{\orbm\orbn}&=\mathscr{G}'^{\orbm\orbn},
\end{align*}
where
\begin{align*}
\mathscr{G}'^{\orbm\orbn}=G'^{\redA\redB}(D'_\redA X^\orbm  )D'_\redB X^\orbn
\end{align*}
has been defined, which is the spacetime component of $G'^{\redA\redB}$.

In order to derive the equation of motion for $G$ membranes,
we need the expression for the extrinsic curvature vector:
  \begin{align*}
K^\sta_{\wsB\wsC}&=D_\wsB D_\wsC X^\sta+\Gamma^\sta_{\stb\stc}(D_\wsB X^\stb)D_\wsC X^\stc\\
&=X^\sta_{,\wsB\wsC}-{}^G\Gamma^\wsA_{\wsB\wsC} X^\sta{}_{,\wsA}
+\Gamma^\sta_{\stb\stc}X^\stb{}_{,\wsB}X^\stc{}_{,\wsC},
  \end{align*}
where ${}^G\Gamma^\wsA_{\wsB\wsC}$ denotes the Christoffel symbol 
with respect to $G_{\wsA\wsB}$.

Noting that
\begin{align*}
  {}^G\Gamma^\wsA_{\wsB\wsC}=G^{\wsA\wsD}X^\sta{}_{,D}g_{\sta\std}
(X^\std{}_{,\wsB\wsC}+\Gamma^\std_{\stb\stc}X^\stb{}_{,\wsB}X^\stc{}_{,\wsC}),
\end{align*}
we have another expression for the extrinsic curvature vector
\begin{align*}
  K^\sta_{\wsA\wsB}
=(\delta^\sta{}_\std-\mathscr{G}^\sta{}_\std)
(X^\std{}_{,\wsA\wsB}+\Gamma^\std_{\stb\stc}X^\stb{}_{,\wsA}X^\stc{}_{,\wsB}).
\end{align*}
  The direct computations show
\begin{align}
K^\orbm_{\homi\homj}&=
N'^{\orbm\orbn}
\left(f_{\kilk(\kili\kilj)}w^\kilk{}_\orbn-\dfrac{1}{2}\phi_{\kili\kilj,\orbn}\right)
\sigma^\kili{}_\homi\sigma^\kilj{}_\homj,
\label{eq:Kmij}\\
K^\orbm_{\homi\redB}&=
\dfrac{1}{2}N'^{\orbm\orbn}
\left(f_{\kili\kilj\kilk}w^\kilj{}_\orbn w^\kilk{}_\orbl
+\phi_{\kili\kilj,\orbn}w^\kilj{}_\orbl
-2\phi_{\kili\kilj}w^\kili{}_{[\orbn,\orbl]}\right)
\sigma^\kili{}_\homi D'_{\redB}X^\orbl,
\label{eq:KmiB}\\
\label{eq:KmAB}K^\orbm_{\redA\redB}&=K'^\orbm_{\redA\redB}\\
&+N'^{\orbm\orbn}
\left(-\dfrac{1}{2}\phi_{\kili\kilj,\orbn}w^\kili{}_\orbl w^\kilj{}_\orbr
+2\phi_{\kili\kilj}w^\kili{}_{[\orbn,\orbl]}w^\kilj{}_\orbr\right)
(D'_{(\redA} X^\orbl)D'_{\redB)} X^\orbr,
\nonumber\\
K^\homi_{\wsA\wsB}&=\sigma_\kili{}^\homi w^\kili{}_\orbm K^\orbm_{\wsA\wsB},
\label{eq:KiAB}
\end{align}
where $D'_\redA$ denotes the covariant derivative compatible with
$G'_{\redA\redB}$; 
$N'^{\orbm\orbn}$ the projection onto the normal space to $W/G$ 
in $M/G$, defined as
\begin{align*}
  N'^{\orbm\orbn}=h^{\orbm\orbn}-\mathscr{G}'^{\orbm\orbn};
\end{align*}
and
\begin{align*}
K'^\orbm_{\redA\redB}=
D'_\redA D'_\redB X^\orbm
+{}^h\Gamma^\orbm_{\orbn\orbl}(D'_\redA X^\orbn)D_\redB X^\orbl
\end{align*}
is the extrinsic curvature vector
of $W/G$ relative to $(M/G,h)$.

Then, the equation of motion is calculated as
\begin{align*}
  K^\orbm&=K'^\orbm
+N'^{\orbm\orbn}\left(
f_{\kili\kilj}{}^\kilj w^\kili{}_\orbn-\dfrac{1}{2}\phi^{-1}\phi_{,\orbn}
\right)
=0,
\end{align*}
where $K'^\orbm=G'^{\redA\redB}K'^\orbm_{\redA\redB}$ is the
mean curvature vector of $W/G$ relative to $(M/G,h)$, and
  we abbreviate as $\phi=\operatorname{det}\phi_{\kili\kilj}$.
The remaining equation $K^\homi=0$ does not give further restriction
since
\begin{align*}
K^\homi=\sigma_\kili{}^\homi w^\kili{}_\orbm K^\orbm
\end{align*}
holds.

This resembles the equation of motion for Nambu-Goto membranes,
but with the extra force term. 
We can partially reduce the force term via the conformal transformation
\begin{align*}
  h_{\orbm\orbn}&=|\phi|^{-1/\dimrw}\widetilde h_{\orbm\orbn},\\
G'_{\redA\redB}&=|\phi|^{-1/\dimrw}\widetilde G_{\redA\redB},
\end{align*}
where $\dimrw=\dimw-\dimg$.
The inverses of $\widetilde h_{\orbm\orbn}$ and
$\widetilde G_{\redA\redB}$ are, respectively, written
as $\widetilde h^{\orbm\orbn}$ and
$\widetilde G^{\redA\redB}$.

The extrinsic curvature vector of $W/G$ relative to
$(M/G,\widetilde h)$ is written as
\begin{align*}
    \widetilde K^\orbm_{\redA\redB}
=\widetilde D_\redA\widetilde D_\redB X^\orbm
+\widetilde \Gamma^\orbm_{\orbn\orbl}(\widetilde D_\redA X^\orbn)
\widetilde D_\redB X^\orbl,
\end{align*}
where $\widetilde D_{\redA}$ denotes the covariant derivative
with respect to the conformally transformed world sheet metric
$\widetilde G_{\redA\redB}$ and 
$\widetilde{\Gamma}^\orbm_{\orbn\orbl}$ the Christoffel symbol with respect to
$\widetilde G_{\redA\redB}$.
Here and in what follows, 
the indices for conformally transformed quantities
are raised or lowered in terms of
$\widetilde G^{\redA \redB}$,
$\widetilde G_{\redA \redB}$,
$\widetilde h^{\orbm\orbn}$,
and
$\widetilde h_{\orbm\orbn}$.

The extrinsic curvature vector undergoes the conformal transformation
as
\begin{align*}
  \widetilde K^\orbm_{\redA\redB}
  = K'^\orbm_{\redA\redB}-\dfrac{1}{2\dimrw}
\phi^{-1}\phi_{,\orbn}
\widetilde N^{\orbm\orbn}
\widetilde G_{\redA\redB},
\end{align*}
where the projection tensor $\widetilde N^{\orbm\orbn}$ has been
 defined as
 \begin{align*}
   \widetilde N^{\orbm\orbn}=|\phi|^{-1/\dimrw}N'^{\orbm\orbn}.
 \end{align*}


Finally the equation of motion for $G$ membranes becomes
\begin{align}
\widetilde K^\orbm
=\widetilde N^{\orbm\orbn}f_{\kilj\kili}{}^\kilj w^\kili{}_{\orbn},
\label{eq:eom-tilK}
\end{align}
where
\begin{align*}
\widetilde K^\orbm&=
\widetilde G^{\redA\redB}\widetilde K^\orbm_{\redA\redB}
=\widetilde D_\redC\widetilde D^\redC X^\orbm
+\widetilde \Gamma^\orbm_{\orbn\orbl}
(\widetilde D_\redC X^\orbn)
\widetilde D^\redC X^\orbl
\end{align*}
is the mean curvature vector of $W/G$ relative to
the orbit space $(M/G,\widetilde h)$.
In this way,  the force term generally appears at the right-hand side
of the reduced equation of motion (\ref{eq:eom-tilK}).

In Ref.~\cite{KKI15}, 
the $G$-invariant Nambu-Goto membranes are considered in the case of
$\dimrw=1$, and 
it is argued that 
the equation of motion 
reduces to 
the geodesic equation in the conformally
transformed orbit space $(W/G,\widetilde h)$, 
which is  based on the dimensional reduction at the action  level,
\begin{align*}
  S&=-\tau\int_W d^\dimw s\sqrt{|G|}
=-\tau\int_G d^\dimg \alpha|\operatorname{det} \sigma^\kili{}_\homi|
\int_{ W/G}d \beta^1 \sqrt{|\widetilde G_{11}|}\\
&\propto\int_{ W/G}d \beta^1 \sqrt{|\widetilde G_{11}|}.
\end{align*}
The last expression gives the geodesic action.
However it turns out that 
  it generally does not  produce a correct equation of motion
due to the presence of the force term,
as we have explicitly shown.

In a certain special cases, the force term becomes zero so that the 
the configuration of the
$G$-invariant membranes corresponds to
the extremal surface in the orbit space $(W/G,\widetilde h)$,
or to the geodesic when $\dimrw=1$.
They include when
\begin{enumerate}
\renewcommand{\labelenumi}{(\roman{enumi})}
\item $G$ is Abelian: 
All the structure constants $f_{\kili\kilj}{}^\kilk$ become zero. 
This includes the case when
the orbit $Gx$ is one dimensional.
\item $G$ is semisimple and compact:
The Jacobi identity for the structure constants implies that
$f_{\kili\kilj}{}^\kilj=0$ automatically holds.
\item Orthogonal distributions of $G$-orbits are integrable:
When the orbit $Gx$ is everywhere orthogonal to the orbit space, $w^\kili{}_\homi$
becomes identically zero.
\end{enumerate}

\section{$G$ membranes coupled to scalar map}
\label{sec:4}
 It would be natural to ask whether 
a reduction mechanism similar to that
shown in the previous section works in the presence of the external fields. 
As a simple case, 
  we here consider
the membranes coupled to a single complex scalar field without $U_1$ gauge couplings. 

Assume that there is a complex scalar field $\psi:W\longrightarrow \boldsymbol{C}$ on the membrane.
We consider the following model for the membrane coupled to a scalar map:
\begin{align*}
S[X^\sta,\psi]&=S_{\rm NG}[X^\sta]+S_{\psi}[X^\sta,\psi],\\
S_{\rm NG}[X^\sta]&=-\tau\int_W ds^1\dots ds^\dimw
\sqrt{|G|},\\
S_\psi[X^\sta,\psi]&=-\kappa\int_W ds^1\dots ds^\dimw
\sqrt{|G|}\left[
G^{\wsA\wsB}(D_\wsA\psi^*)D_\wsB\psi+U(\psi^*\psi)
\right].
\end{align*}

The first variation of this action with respect to $X^\sta$ gives
the equation of motion for the membrane
\begin{align}
\dfrac{1}{\sqrt{|G|}}g^{ab}\dfrac{  \delta S}{\delta X^b}
&=\dfrac{1}{\sqrt{|G|}}\partial_{\wsA}\left[
  \sqrt{|G|}(\tau G^{\wsA\wsB}-T^{\wsA\wsB})D_BX^a
  \label{eq:EL-2-1}
\right]\\
&+\Gamma^a_{bc}\left[\tau \mathscr{G}^{bc}-
T^{\wsB\wsC}(D_\wsB X^b)D_\wsC X^c
\right]=0,\nonumber
\end{align}
where the stress-energy tensor 
\begin{align*}
  T^{\wsA\wsB}&=\kappa[2(D^{(\wsA}\psi^*)D^{\wsB)}\psi
-(D_{\wsC}\psi^*)(D^{\wsC}\psi) G^{\wsA\wsB}
-UG^{\wsA\wsB}]
\end{align*}
on the world sheet $W$ has been defined.

The first variation with respect to $\psi^*$ gives the
wave equation
\begin{align}
\label{eq:ELS-2-1}  \dfrac{1}{\kappa\sqrt{|G|}}  \dfrac{\delta S}{\delta \psi^*}=
    D_\wsC D^\wsC \psi-U'\psi=0
\end{align}
for $\psi$.
This implies the local conservation law for the energy
\begin{align*}
  D_\wsA T^{\wsA\wsB}=0.
\end{align*}
Using this equation, Eq.~(\ref{eq:EL-2-1}) reduces to
\begin{align}
  \tau K^\sta -T^{\wsA\wsB}K^\sta_{\wsA\wsB}=0,
\label{eq:EL-2-2}
\end{align}
in terms of the extrinsic curvature vector.

Here we assume the $G$-invariant configuration
for the metric
\begin{align*}
  g=\phi_{\kili\kilj}
(\sigma^\kili-w^\kili{}_\orbm dz^\orbm)
(\sigma^\kilj-w^\kilj{}_\orbn dz^\orbn)
+|\phi|^{-1/\dimrw}\widetilde h_{\orbm\orbn}dz^\orbm dz^\orbn,
\end{align*}
as in the previous section, and for the membrane and the
scalar field on it:
\begin{align*}
  X^\homi=\alpha^\homi,~~~
X^\orbm=X^\orbm(\beta^\redA),~~~
\psi=\psi(\beta^\redA).
\end{align*}

Then, Eq. (\ref{eq:EL-2-2}) reduces to
\begin{align*}
\tau K^\orbm-T^{\wsA \wsB}K^\orbm_{\wsA\wsB}
&=|\phi|^{1/\dimrw}\biggl\{
\tau \widetilde K^\orbm-\widetilde T^{\redA\redB}\widetilde K^\orbm_{\redA\redB}\\
&\nonumber+\widetilde N^{\orbm\orbn}
\biggl[
\tau f_{\kili\kilj}{}^\kilj w^\kili{}_\orbn\\
&+|\phi|^{1/\dimrw}(\widetilde D_\redC\psi^*)(\widetilde D^\redC\psi)
\left(f_{\kili\kilj}{}^\kilj w^\kili{}_\orbn-
\dfrac{1}{\dimrw}\phi^{-1}\phi_{,\orbn}
\right)
\biggr]
\biggr\}=0,
\end{align*}
and
\begin{align*}
  \tau K^\homi-T^{\wsA \wsB}K^\homi_{\wsA\wsB}
=\sigma_\kili{}^\homi w^\kili{}_\orbm(\tau K^\orbm-T^{\wsA \wsB}K^\orbm_{\wsA\wsB})=0,
\end{align*}
where the reduced stress-energy tensor is defined by
\begin{align*}
  \widetilde T^{\redA\redB}
  &=\kappa\left\{
  |\phi|^{1/\dimrw}
  \left[
    2(\widetilde D^{(\redA}\psi^*)\widetilde D^{\redB)}\psi
               -(\widetilde D_{\redC}\psi^*)(\widetilde D^{\redC}\psi) \widetilde G^{\redA\redB}
             \right]
  -U\widetilde G^{\redA\redB}
  \right\}.
\end{align*}
On the other hand, the wave equation (\ref{eq:ELS-2-1})
becomes
\begin{align}
  \widetilde D_\redC (|\phi|^{1/\dimrw}\widetilde D^\redC \psi)
- U'\psi=0.
\label{eq:ELS-2-2}
\end{align}
In summary, the equation of motion reduces to
\begin{align}
\tau \widetilde K^\orbm-\widetilde T^{\redA\redB}\widetilde K^\orbm_{\redA\redB}
&\label{eq:EL-2-3}
=\widetilde N^{\orbm\orbn}
\biggl[
\tau f_{\kilj\kili}{}^\kilj w^\kili{}_\orbn\\
&\nonumber+|\phi|^{1/\dimrw}(\widetilde D_\redC\psi^*)(\widetilde D^\redC\psi)
\left(f_{\kilj\kili}{}^\kilj w^\kili{}_\orbn+
\dfrac{1}{\dimrw}\phi^{-1}\phi_{,\orbn}
\right)
\biggr].
\end{align}
Except for the term with the factor $f_{\kilj\kili}{}^\kilj w^\kili{}_\orbn$,
Eqs.~(\ref{eq:EL-2-2}) and (\ref{eq:EL-2-3}) are derived from the action
\begin{align}
 \widetilde S[X^\orbm,\psi]=-\int_{W/G}d\beta^1\dots d\beta^\dimrw
\sqrt{|\widetilde G|}
\left\{
  \tau+\kappa\left[
  |\phi|^{1/\dimrw}(\widetilde D_\redC\psi^*)
  \widetilde D^\redC\psi+U(\psi^*\psi)
  \right]
\right\},
\end{align}
obtained via the naive dimensional reduction.

\section{Coupling to differential form field}
\label{sec:5}
As another model for matter coupling, we consider a background differential form field $\omega$,
which is a $\dimw$-form field on $M$.

The simplest model would be given by
\begin{align*}
S[X^\sta]&=S_{\rm NG}[X^\sta]+S_\omega[X^\sta],\\
S_{\rm NG}[X^\sta]&=-\tau\int_W ds^1\dots ds^\dimw \sqrt{|G|},\\
S_\omega[X^\sta]&=-\lambda\int_W i^*\omega\\
&=\dfrac{\lambda}{\dimw !}\int_W ds^1\dots ds^\dimw
\omega_{\sta_1\dots\sta_\dimw} (D_{\wsA_1}X^{\sta_1})\dots (D_{\wsA_\dimw} X^{\sta_\dimw})
\epsilon^{\wsA_1\dots\wsA_\dimw},
\end{align*}
where $\epsilon^{\wsA_1\dots\wsA_\dimw}$ denotes the $\dimw$-index
Levi-Civita symbol on $W$ such that $\epsilon^{12\dots\dimw}=-1$.

The first variation of the action is calculated as
\begin{align*}
  \dfrac{\delta S}{\delta X^a}
&=\tau \sqrt{|G|} g_{\sta\stb}K^\stb
+\dfrac{\lambda (\dimw+1)}{\dimw !}
\omega_{[\sta_1\dots\sta_\dimw,\sta]}
(D_{\wsA_1}X^{\sta_1})\dots (D_{\wsA_\dimw}X^{\sta_\dimw})
\epsilon^{\wsA_1\dots\wsA_\dimw}.
\end{align*}
Thus the equation of motion for the membrane becomes
\begin{align}
  \tau  K^\sta
=-\dfrac{\lambda (\dimw+1)}{\dimw !\sqrt{|G|}}
g^{\sta\stb}\omega_{[\sta_1\dots\sta_\dimw,\stb]}
(D_{\wsA_1}X^{\sta_1})\dots (D_{\wsA_\dimw}X^{\sta_\dimw})
\epsilon^{\wsA_1\dots\wsA_\dimw}.
\label{eq:EL-4-1}
\end{align}

We assume that both the membrane and the background $\dimw$-form field
are $G$-invariant.
The general form of the $G$-invariant $\dimw$-form is 
\begin{align*}
 (\mbox{left invariant $p$-form on $Gx$})\wedge [\mbox{($\dimw-p$)-form on
$M/G$}]
\end{align*}
or their linear combination.
Among these, only the $p=\dimg$ case 
  results in the reduction of the system
to the membrane equation in $M/G$.
Hence we choose
\begin{align*}
  \omega=\sigma^1\wedge\sigma^2\wedge\dots\wedge \sigma^\dimg\wedge
\widetilde\omega,
\end{align*}
where $\widetilde\omega$ is the $\dimrw$-form on $M/G$.
In terms of coordinate components, we assume that
\begin{align*}
  X^\homi=\alpha^\homi,~~
X^\orbm=X^\orbm(\beta^\redA),\\
\omega_{\sta_1\dots\sta_\dimw}
=\dfrac{\dimw!}{\dimrw!}
\sigma^1{}_{[\sta_1}\dots\sigma^\dimg{}_{\sta_\dimg}
\widetilde\omega(x^{\orbm})_{\sta_{\dimg+1}\dots\sta_{\dimg+\dimrw}]}.
\end{align*}

Then, Eq.~(\ref{eq:EL-4-1}) reduces to
the equation for the membrane in $M/G$ as
  \begin{align}
 \tau  
\widetilde K^\orbm&=(-1)^{s+t}
\dfrac{\lambda (\widetilde{\dimw}+1)}{\widetilde{\dimw} !\sqrt{|\widetilde {G}|}}
\widetilde h^{\orbm\orbn}
\widetilde\omega_{[\orbm_1\dots\orbm_{\widetilde{\dimw}},\orbn]}
(D_{\redA_1}X^{\orbm_1})\dots 
(D_{\redA_{\widetilde{\dimw}}}X^{\orbm_{\widetilde{\dimw}}})
\epsilon^{\redA_1\dots\redA_{\widetilde{\dimw}}}
\label{eq:EL-4-2}\\
&\nonumber+\tau\widetilde N^{\orbm\orbn}f_{\kilj\kili}{}^\kilj w^\kili{}_\orbn
  \end{align}
The  factor $(-1)^s$ in the first term on the rhs
is $(+1)$ if $\widetilde G_{\redA\redB}$ has Riemannian signature and  $(-1)$
if Lorentzian,
and the factor $(-1)^t$ denotes the signature of $\det{\sigma^\kili{}_\homi}$.
The $\dimrw$-index Levi-Civita symbol on $W/G$ 
has been normalized such that $\epsilon^{\dimg+1,\dimg+2,\dots,\dimg+\dimrw}
=(-1)^s$.

Except for the 
final force term,
Eq.~(\ref{eq:EL-4-2})
has the same form as Eq.~(\ref{eq:EL-4-1}), which is derived from
the naive reduced action
\begin{align}
 \widetilde S[X^\orbm,\widetilde \omega]
  =-\tau \int_{W/G}d\beta^1\dots d\beta^\dimrw\sqrt{|\widetilde G|}
  -(-1)^t\lambda\int_{W/G}i^*\widetilde\omega.
    \end{align}

\section{Concluding remarks}
\label{sec:6}
We have considered  in general settings the motion of test membranes
on which the group $G$ of 
spacetime isometries acts.
We have found that the configuration of Nambu-Goto membranes
is described by the Nambu-Goto membranes in a quotient manifold
with the appropriate projected metric,
  if at least one of the following conditions holds;
(i) $G$ is Abelian,
(ii) $G$ is semisimple and compact,
or (iii) the orthogonal distribution
of the orbit of $G$ is integrable.
We have also obtained  similar results
for the membranes
 coupled with the scalar maps or the differential form fields.

  At the same time, it should be emphasized that 
the usual dimensional reduction procedure
at the action level is not always justified.
This is because the variational principle for dimensionally reduced action does not 
incorporate the variation of membranes with inhomogeneous variation
with respect to the $G$-orbits.

  Nevertheless, the correct equation of motion for $G$-membranes
derived here is only slightly different from the naive equation of motion
by  force terms written with local geometrical quantities.
Hence,
 our formalism would be useful when
we seek for more general string/membrane
solutions in spacetimes with isometries,
and when we classify such solutions.

\section*{Acknowledgments} 
We thank Professor Hideki Ishihara for fruitful discussion and suggestions.

\end{document}